\documentclass{article}

\usepackage{arxiv}

\usepackage[utf8]{inputenc} 
\usepackage[T1]{fontenc}    
\usepackage{hyperref}       
\usepackage{url}            
\usepackage{booktabs}       
\usepackage{amsfonts}       
\usepackage{nicefrac}       
\usepackage{microtype}      
\usepackage{lipsum}
\usepackage{graphicx}
\graphicspath{ {./images/} }
\usepackage{amsmath}
\usepackage{amssymb}

\title{Strong Electron Correlation from Partition Density Functional Theory}

\author{
 Yi Shi \\
  Department of Chemistry\\
  Purdue University\\
  West Lafayette, IN 47906 \\
  \texttt{shi427@purdue.edu} \\
   \And
 Yuming Shi \\
  Department of Physics and Astronomy\\
  Purdue University\\
  West Lafayette, IN 47906 \\
  \texttt{shi449@purdue.edu} \\
  \And
 Adam Wasserman \\
  Department of Chemistry\\
  Purdue University\\
  West Lafayette, IN 47906 \\
  \texttt{awasser@purdue.edu} \\
}

\begin{document}
\maketitle
\begin{abstract}
Standard approximations for the exchange-correlation (XC) functional in Kohn-Sham density functional theory (KS-DFT) typically lead to unacceptably large errors when applied to strongly-correlated electronic systems. Partition-DFT (PDFT) is a formally exact reformulation of KS-DFT in which the ground-state density and energy of a system are obtained through self-consistent calculations on isolated fragments, with a partition energy representing the \textit{inter}-fragment interactions. Here we show how typical errors of the local density approximation (LDA) in KS-DFT can be largely suppressed through a simple approximation, the generalized overlap approximation (GOA), for the partition energy in PDFT. Our method is illustrated on simple models of one-dimensional strongly-correlated linear hydrogen chains. The GOA, when used in combination with the LDA for the fragments, improves the LDA dissociation curves of hydrogen chains and produces results that are comparable to those of spin-unrestricted LDA, but without breaking the spin symmetry. GOA also induces a correction to the LDA electron density that partially captures the correct density dimerization in strongly-correlated hydrogen chains. Moreover, with an additional correction to the partition energy, the approximation is shown to produce dissociation energies in quantitative agreement to calculations based on the Density Matrix Renormalization Group method.
\end{abstract}


\section{\label{sec:introduction}Introduction}
Perdew's Jacob's ladder of approximations to the exchange-correlation (XC) energy functional $E_{\rm xc}[n]$ \cite{perdew2001jacob} provides a map that guides us toward the goal of finding usefully accurate functionals for Kohn-Sham \cite{KS1965} and Generalized Kohn-Sham \cite{seidl1996generalized} Density Functional Theory (DFT).   One of the greatest challenges at every step of the ladder is the description of strongly-correlated electronic systems. The local density approximation (LDA), the first density functional approximation (DFA) to the XC functional, typically fails when applied to strongly-correlated systems.
A simple, representative case is that of a closed-shell molecule stretched beyond its equilibrium bond length, when its ground state wavefunction cannot be accurately represented by a single Slater determinant. Near-degeneracies present at the stretched confirguration lead to large static-correlation errors in LDA calculations \cite{SCError1,SCError2,SCError3}. The LDA may yield quantitatively correct energies in the weakly-correlated region near the equilibrium bond length, but overestimates the energy significantly as the molecule is stretched. 
DFAs at higher rungs of the ladder, including the generalized gradient approximation (GGA)\cite{PW91,PBE}, meta-GGA\cite{TPSS,SCAN}, and hybrid functionals\cite{B3LYP1,B3LYP2,PBE0}, suffer from similar problems. Those DFAs do account for a certain fraction of correlation effects, but they typically continue to fail in the presence of strong electron correlation\cite{Pavarini2021,Perdew2002}. 

Strong electron correlation plays a central role in many exotic properties of condensed matter, such as high-temperature superconductivity\cite{SuperCon1,SuperCon2}, quantum Hall effects\cite{QHall1,QHall2}, and Mott-type metal-insulator transitions\cite{MottTrans}. Various methods that lie outside of the KS-DFT framework have been developed for treating such systems. Some of these lead to quantitatively correct results by adding corrections to one-electron theories. For instance, in the popular  DFT+$U$ method \cite{Anisimov1991, Liechtenstein1995}, a somewhat \textit{ad hoc} parameter, Hubbard $U$,  can be imposed on certain states of the system to fix the errors caused by XC functionals in KS-DFT. Many-body methods where the one-electron approximation is not applied, such as the dynamical mean field theory (DMFT)\cite{DMFT1,DMFT2,DMFT3} and the density matrix renormalization group (DMRG)\cite{DMRG1,DMRG2,DMRG3} method, are also powerful tools for strongly correlated systems, but they are typically much more computationally expensive than DFT.

Linear chains of hydrogen atoms are among the simplest models that can reveal the effects of strong electron correlation. They are computationally tractable and have been well investigated by researchers from diverse areas\cite{HChain1,HChain2,HChain3,HChain4,HChain5,Stoudenmire2012,Motta2017,Motta2020}. Despite their simplicity,  linear hydrogen chains embody richer chemical complexities than simpler models such as the Hubbard model, so they can be used as step stones toward realistic materials of the strongly-correlated type. In particular, the hydrogen chains display characteristics of strongly-correlated physics when the interatomic distances increase beyond equilibrium separations. Here we study simplified models in one dimension (1D) that retain some of the essential properties of their three-dimensional counterparts.
We study 1D strongly correlated hydrogen chains using a fragment-based DFT approach, Partition-DFT (PDFT)\cite{PDFT1,PDFT2}, and demonstrate that a simple approximation to the partition energy functional of PDFT \cite{OA} captures key signatures of strongly-correlated physics, as demonstrated through detailed comparisons with numerically exact DMRG calculations \cite{Stoudenmire2012}. 

We begin by summarizing the essential background of Partition-DFT in Sec.II, where we also introduce an extension of a recent Overlap Approximation \cite{OA} to the partition energy (GOA). After  providing computational details in Sec.III, we illustrate in Sec.IV how restricted LDA dissociation curves of 1D Hydrogen chains are corrected by
the GOA toward the {\em unrestricted} LDA energies but without breaking the correct spin symmetry. Another key signature of strongly-correlated physics, the dimerization of electron density, is discussed in Sec.V, where we demonstrate numerically that the GOA provides a dimerization measure lying roughly in between the incorrect LDA values and those of exact DMRG calculations.  Finally, we discuss in Sec.VI a possible correction to the GOA that brings LDA energies even closer to exact DMRG results.

\section{\label{sec:PDFT}Partition-DFT}

PDFT \cite{PDFT1,PDFT2} is a formally exact reformulation of KS-DFT in which a system of interacting fragments is mapped onto one of {\em non-interacting} fragments with the same total density. The ground state energies $\{E_{\alpha}\}$ and densities $\{n_{\alpha}\}$ of these non-interacting fragments are obtained through self-consistent KS-DFT calculations that minimize the sum of fragment energies $E_f = \sum_{\alpha}E_{\alpha}[n_{\alpha}]$ under the constraint: 
\begin{equation}
n_f(\textbf{r}) = \sum_{\alpha} n_{\alpha} (\textbf{r}) = n_M (\textbf{r}),
\label{eq:nf}
\end{equation}
ensuring that the sum of fragment densities $n_f(\bf r)$ and the true ground state density of the entire system $n_M (\textbf{r})$ are equal. A one-body local potential enforcing the constraint of Eq.~(\ref{eq:nf}), the partition potential $v_p(\textbf{r})$, acts as the Lagrange multiplier in this minimization. The total energy in PDFT is calculated as
\begin{equation}
E[\{n_{\alpha}\}] = E_f [\{n_{\alpha}\}] + E_p [\{n_{\alpha}\}],
\label{eq:Ef+Ep}
\end{equation}
where the partition energy $E_p$ is the contribution from $v_p({\bf r})$ to the total energy due to its presence in the fragment KS equations, accounting for the interaction between fragments. The partition potential $v_p({\bf r})$ is given by the functional derivative 
$v_p(\textbf{r}) = \delta E_p[\{n_{\alpha}\}]/\delta n_{\alpha}(\textbf{r})$ at the minimizing densities. The partition energy in PDFT can be decomposed into four non-additive KS components:
\begin{equation}
\begin{split}
    E_p[\{n_{\alpha}\}] = &T_s^{\rm nad}[\{n_{\alpha}\}] + E_{\rm ext}^{\rm nad}[\{n_{\alpha}\}] \\
    &+ E_{\rm H}^{\rm nad}[\{n_{\alpha}\}] + E_{\rm xc}^{\rm nad}[\{n_{\alpha}\}],
\end{split}
\label{eq:Epdecompose}
\end{equation}
where, for instance, 
$E_{\rm xc}^{\rm nad}[\{n_{\alpha}\}] = E_{\rm xc}[n_f] - \sum_{\alpha}E_{\rm XC}[n_{\alpha}]$.

With $E_f$ and $E_p$ accounting for \textit{intra}- and \textit{inter}-fragment interactions, respectively, PDFT is in principle exactly equivalent to KS-DFT, that is, it produces the same results as KS-DFT for a given DFA (see for example Fig.\ref{fig:H2KSvsP}). However, PDFT offers additional flexibility as one can construct unique approximations from fragment properties, making it possible to go beyond approximate KS-DFT. In particular, 
in the \textit{overlap approximation} (OA) for binary fragmentation \cite{OA,Yuming2023}, the partition energy of Eq.(\ref{eq:Epdecompose}) is approximated by multiplying its XC-component by an overlap functional $S^{\rm OA}[\{n_\alpha\}]$ of the fragment densities:
\begin{equation}
    E_p^{\rm OA} = T_s^{\rm nad} + E_{\rm ext}^{\rm nad}+E_{\rm H}^{\rm nad}+S^{\rm OA}E_{\rm XC}^{\rm nad}.
\label{eq:EpOA}
\end{equation}
For a system partitioned into two fragments \textit{A} and \textit{B}, defining the overlap as
\begin{equation}
    S^{\rm OA}[n_A,n_B] = {\rm erf}\left\{2 \int \sqrt{n_{A}(\textbf{r})n_{B}(\textbf{r})} d\textbf{r} \right\},
\label{eq:SOA}
\end{equation}
has been shown to yield quantitatively correct LDA and GGA dissociation curves for singly-bonded diatomic molecules\cite{OA}. The explanation is simple and physically sound: When applied in a spin-restricted manner, standard DFAs like LDA typically lead to a static correlation error for stretched molecules. Consider $\rm H_2$: The molecule remains a spin-singlet in the dissociation limit and, with the exact condition that $n_\uparrow({\bf r})=n_\downarrow({\bf r})$, fractional spins must be assigned to the isolated atoms.  As is well known, most  DFAs lead to incorrect energies for such fractional-spin calculations \cite{cohen2008insights}. A spin-unrestricted KS-DFT calculation will improve the description of molecular dissociation beyond the Coulson-Fischer point\cite{CFpoint},  but good energies are obtained at the expense of breaking the symmetry of the spin densities \cite{spinsym1,spinsym2,spinsym3}. The static-correlation error of restricted LDA for stretched H$_2$ is entirely contained in $E_{\rm xc}^{\rm nad}[\{n_\alpha\}]$ \cite{Yuming2023}, and the overlap functional of Eq.(\ref{eq:SOA}), when used within Eq.(\ref{eq:EpOA}), suppresses this error as the molecule is stretched while conserving the correct spin symmetry.  As an error function of the density overlap, $S^{\rm OA}$ has a range between 0 and 1. For molecules with small bond lengths near equilibrium, $S^{\rm OA}$ stays close to 1 and has no noticeable impact on the original $E_p$ in this range of bond lengths. When the system is stretched, however, and it starts experiencing the effects of strong correlation, $S^{\rm OA}$ gradually descends so that the static correlation error inherent in $E_{\rm xc}^{\rm nad}$ is removed without breaking the spin symmetry.

We strive to obtain similar results for hydrogen chains, i.e. improving the energies without breaking the spin symmetry. 
The OA is a malleable approximation in that its form can be tailored to various situations \cite{Yuming2023}.

We define here the \textit{generalized overlap approximation} (GOA) as the expression of Eq.(\ref{eq:EpOA}) but with $S^{\rm OA}$ replaced by:
\begin{equation}
    S^{\rm GOA}[\{n_{\alpha}\}] = {\rm erf}\left\{\frac{2}{N_f-1}\sum_{\langle \alpha,\beta \rangle} \int \sqrt{n_{\alpha}(\textbf{r})n_{\beta}(\textbf{r})} d\textbf{r} \right\},
\label{eq:SGOA}
\end{equation}
where $N_f$ denotes the number of fragments. The symbol $\langle \alpha,\beta \rangle$ means that the sum is over all nearest neighbors among the fragments.  
Eqs.(\ref{eq:EpOA}) and (\ref{eq:SGOA}) will be shown here to not only achieve the goal of correcting the LDA energies without breaking symmetries, but also capturing other key signatures of the strongly-correlated physics that manifests when stretching bonds.

\section{\label{sec:methods}Computational Details}

All calculations are done on 1D hydrogen chains under the Born-Oppenheimer approximation. The nuclei (protons, in this case) are evenly spaced at fixed coordinates in a 1D box with open boundary conditions. The exact ground-state properties of 1D hydrogen chains are computed with DMRG using ITensor\cite{itensor}. Atomic units are used throughout. 

\begin{figure}[b!]
\centering
\includegraphics[scale=0.7]{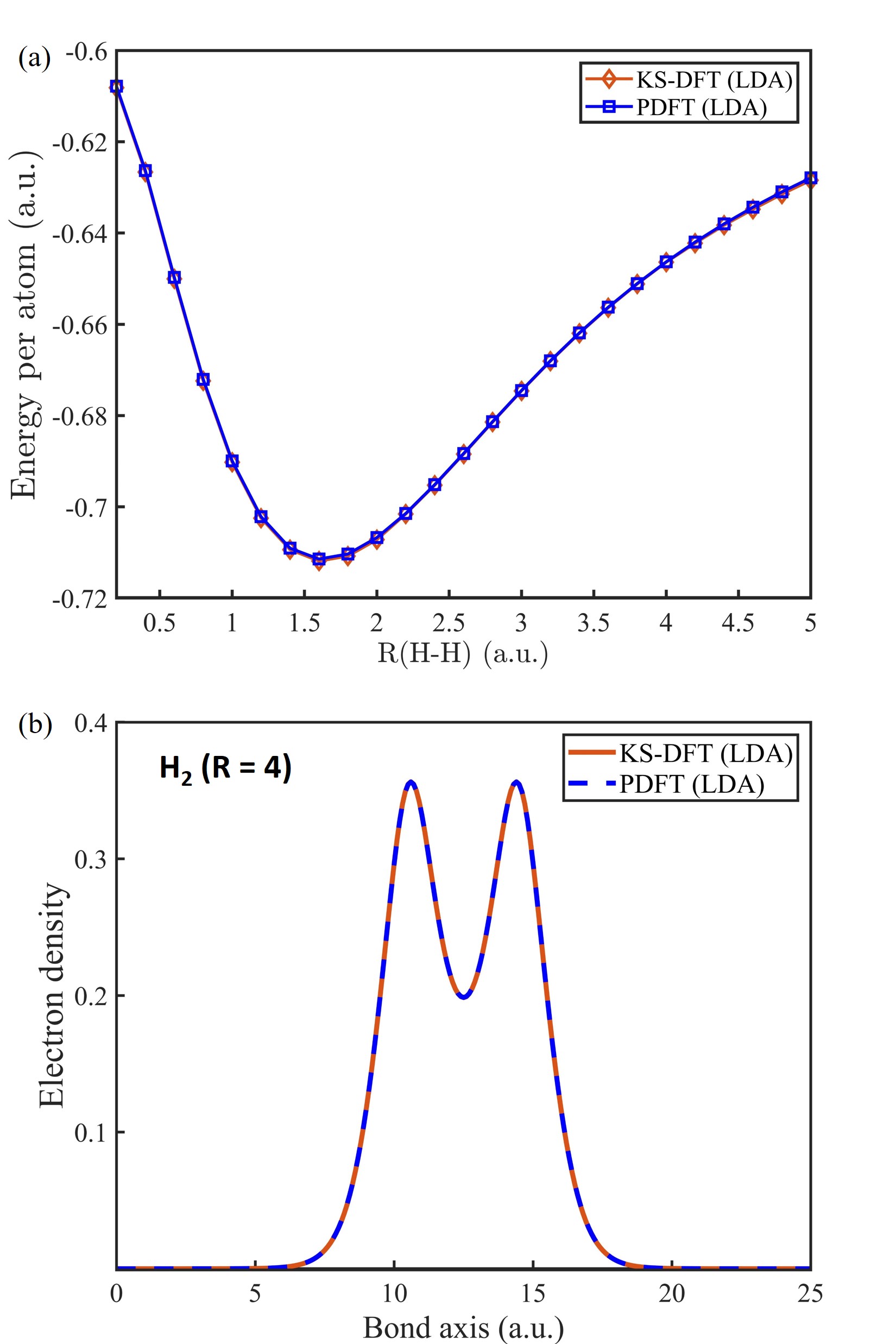}
\caption{\label{fig:H2KSvsP} (a) Dissociation curve (energy per atom) of $\rm H_2$ calculated with KS-DFT (in orange) and PDFT (in blue) in 1D using the same DFA (LDA) for the XC energy functional. (b) Electron density of 1D $\rm H_2$ obtained from KS-DFT (in orange) and PDFT (in blue) calculations using LDA.}
\end{figure}

KS-DFT is implemented in MATLAB on a 1D real-space grid\cite{RGrid,Yan2018} for comparison with DMRG. The kinetic energy in KS-DFT is computed on the grid with a sixth-order finite-difference approximation\cite{RGrid}. The Hartree and external terms are integrated directly on the grid. To avoid the complication resulting from divergences of the bare Coulomb interaction in 1D, the electrostatic interactions are represented by a soft Coulomb potential
\begin{align}
    {\rm Coulomb} \quad &\rightarrow \quad {\rm soft \   Coulomb} \notag\\
    \frac{ZZ'}{\lvert x-x' \rvert}\quad  &\rightarrow \quad \frac{ZZ'}{\sqrt{1+\lvert x-x' \rvert^2}},
\end{align}
where $x$ and $x'$ are the positions of the two particles experiencing the electrostatic interaction with their charges denoted by $Z$ and $Z'$.
LDA is used here for the XC energy. The 1D LDA exchange and correlation functionals are obtained based on the exact exchange and correlation energy of a 1D homogeneous electron gas\cite{LDAEX,LDACO}, respectively.

PDFT is implemented in 1D by self-consistently solving each fragment with 1D KS-DFT. $T_s^{\rm nad}$ in Eq.~(\ref{eq:Epdecompose}) is calculated directly through a density-to-potential inversion\cite{inverse1,inverse2}, and we use LDA again for both the fragment XC energies and $E_{\rm XC}^{\rm nad}$ in PDFT. As mentioned in Sec. \ref{sec:introduction}, the results of KS-DFT are reproduced exactly by PDFT when the same DFA is used in both calculations. This is verified with a numerical calculation on a 1D hydrogen molecule ($\rm H_2$). Fig. \ref{fig:H2KSvsP} shows (a) the dissociation curve (energy per atom as a function of bond length $R$) of $\rm H_2$ and (b) the ground-state electron density of $\rm H_2$ at $R=4$, calculated with KS-DFT and PDFT using LDA as the DFA. In PDFT, the molecule is partitioned as two fragments with one hydrogen atom in each. It is evident from Fig. \ref{fig:H2KSvsP} that both the energy and the density from KS-DFT are exactly reproduced by PDFT.
\clearpage

\section{\label{sec:dissociation}Dissociation Curves}

\begin{figure}[t]
\centering
\includegraphics[scale=0.7]{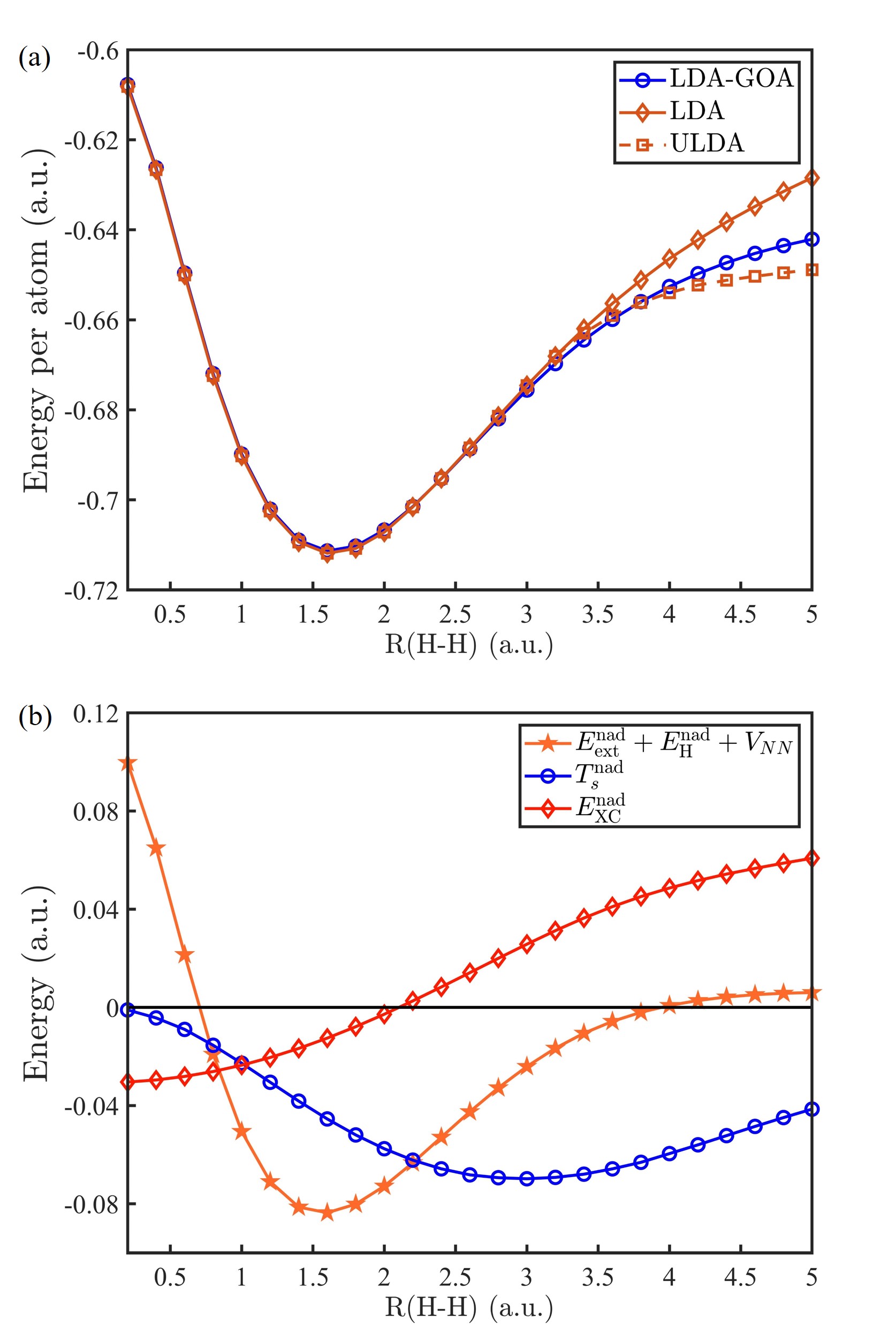}
\caption{\label{fig:H2disso_GOA} (a) Dissociation curve of $\rm H_2$ calculated by combining LDA and GOA (LDA-GOA, blue) in PDFT. LDA (solid orange) and ULDA (dashed orange) curves are obtained through KS-DFT calculations for comparison. (b) Non-additive components of $E_p$, plotted as functions of $R$. LDA is used for $E_{\rm XC}^{\rm nad}$ and GOA is not applied. The repulsion between nuclei $V_{NN}$ is added to $E_{\rm ext}^{\rm nad} + E_{\rm H}^{\rm nad}$ so that they are in the same order of magnitude with $T_s^{\rm nad}$  and $E_{\rm XC}^{\rm nad}$.}
\end{figure}

Analogous to the case of diatomic molecules, a linear chain of atoms experiences effects of strong static correlation as the inter-nuclear separations $R$ are stretched beyond their equilibrium values. 
When standard DFAs are used in KS-DFT to compute the dissociation curves of hydrogen chains, static-correlation errors  emerge in the large $R$ regions.
A density functional method that can be used for strongly correlated systems must, at least, provide a quantitative treatment for these large-$R$ regions of dissociation curves.

\begin{figure}[t]
\centering
\includegraphics[scale=0.7]{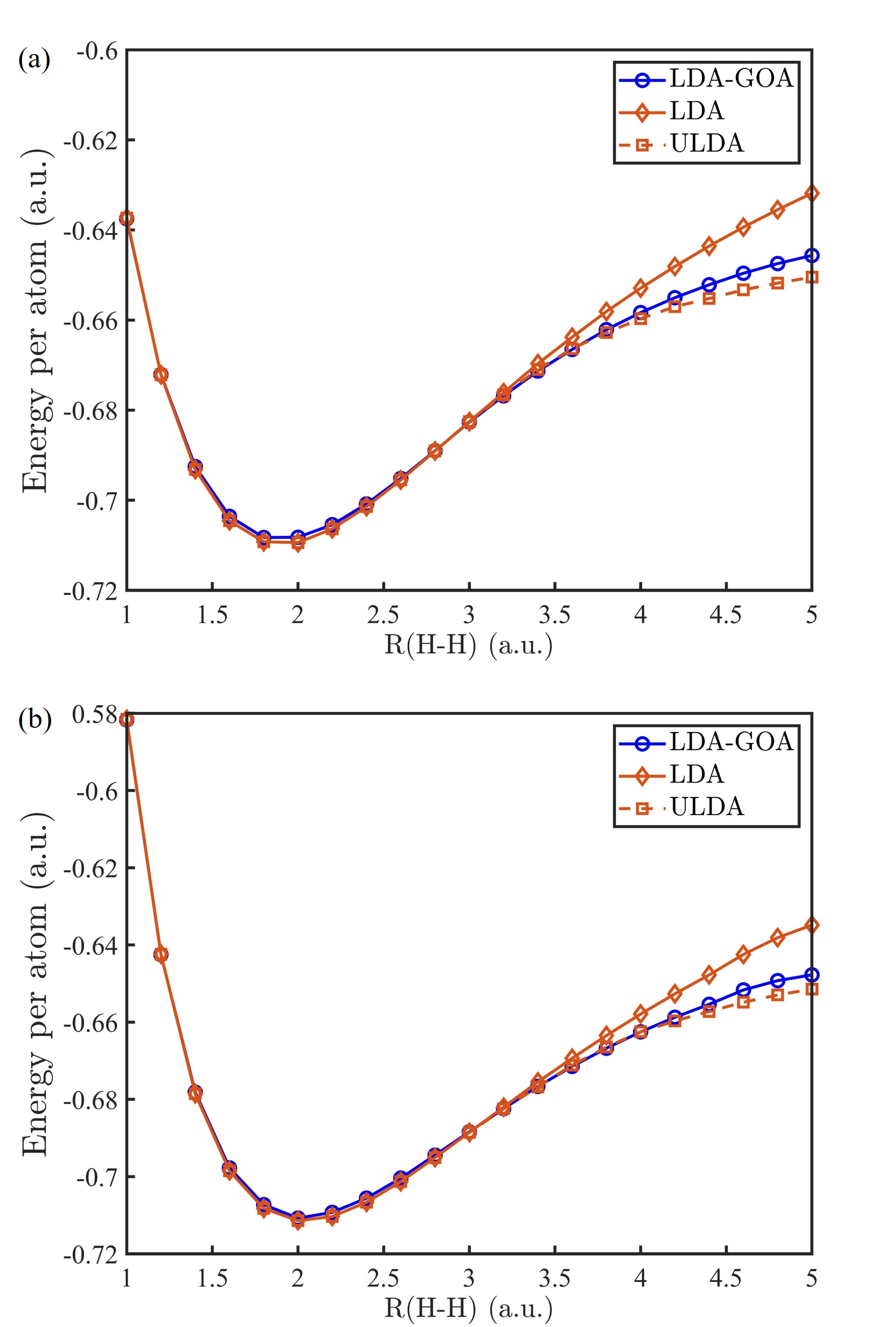}
\caption{\label{fig:H4H10disso_GOA} (a) Dissociation curve of $\rm H_4$ calculated with three methods (LDA, ULDA, and LDA-GOA). (b) Dissociation curve of $\rm H_{10}$ calculated with three methods (LDA, ULDA, and LDA-GOA).}
\end{figure}

We begin by testing the GOA of Eqs.(\ref{eq:EpOA}) and (\ref{eq:SGOA}) on a 1D $\rm H_2$ molecule. For $\rm H_2$ in 3D, the LDA-OA of Eqs.(\ref{eq:EpOA}-\ref{eq:SOA}) has been shown to yield excellent agreement with unrestricted LDA energies, without breaking the spin symmetry \cite{OA}. 
Since the GOA of Eq.(\ref{eq:SGOA}) reduces to the OA of Eq.(\ref{eq:SOA}) for the case of only two fragments, we expect to obtain similar results here, but it is important to check because 1D (with soft Coulomb interactions) differs from 3D.   Fig. \ref{fig:H2disso_GOA}(a) shows the dissociation curve (energy per atom) of 1D $\rm H_2$ calculated by combining GOA and LDA in PDFT, compared with the KS-LDA (solid orange) and unrestricted KS-LDA (ULDA, dashed orange). In the weakly-correlated regime ($R$ near the equilibrium bond length), the three curves are almost the same. Unrestricted KS-LDA does not break the spin symmetry in that range of $R$. $S^{\rm GOA}$ stays close to 1 and thus does not have a significant impact on $E_p$. When the molecule is stretched beyond the Coulson-Fischer point ($R=3.4$a.u.), ULDA produces lower energy by breaking the spin symmetry. As $\rm H_2$ approaches the dissociation limit, $E_{\rm LDA}^{\rm H_2}(R=5)=-1.256$a.u. and $E_{\rm ULDA}^{\rm H_2}(R=5)=-1.298$a.u. The LDA-GOA calculation yields $E_{\rm LDA-GOA}^{\rm H_2}(R=5)=-1.284$a.u. Although the LDA-GOA does not reproduce the ULDA energy at $R=5$a.u. as closely as it does in 3D, it corrects about 70\% of the LDA static-correlation error  (as measured with respect to ULDA) without breaking the spin symmetry. 

Panel (b) of Fig. \ref{fig:H2disso_GOA} shows the four non-additive components of $E_p$ in Eq. (\ref{eq:Epdecompose}) as functions of $R$, all extracted from bare LDA PDFT calculations done without the GOA. As $R \rightarrow \infty$, the only term remaining positive is $E_{\rm XC}^{\rm nad}$, which contains a significant fraction of the static correlation error. The sum of $E_{\rm ext}^{\rm nad}$, $E_{\rm H}^{\rm nad}$, and the inter-nuclear repulsion $V_{NN}$ vanishes in the dissociation limit. The non-additive kinetic energy $T_s^{\rm nad}$ is negative when $R=5$a.u., but it has also been shown to vanish  asymptotically as $R \rightarrow \infty$\cite{Yuming2023}.

We now examine the performance of the GOA on hydrogen chains where, interestingly, the static-correlation error of the LDA is suppressed even more significantly than in H$_2$ as the number of hydrogen atoms increases. Dissociation curves are shown in panels (a) and (b) of Fig.\ref{fig:H4H10disso_GOA} for $\rm H_4$ and $\rm H_{10}$, respectively. Similar to the case of $\rm H_2$, the LDA-GOA energies remain indistinguishable from LDA and ULDA when $R\lesssim 3.6$a.u. ($S^{\rm GOA}\sim 1$ in those regions). For larger separations ($R\gtrsim 3.6$a.u.), $S^{\rm GOA}$ gradually decreases and removes the error contained in $E^{\rm nad}_{\rm xc}$. At $R=5$a.u., the GOA ends up rectifying about 74\% and 78\% of the error caused by the LDA for $\rm H_4$ and $\rm H_{10}$, respectively. 

\clearpage
\section{\label{sec:dimerization}Dimerization of Electron Density}

\begin{figure}[b]
\centering
\includegraphics[scale=0.3]{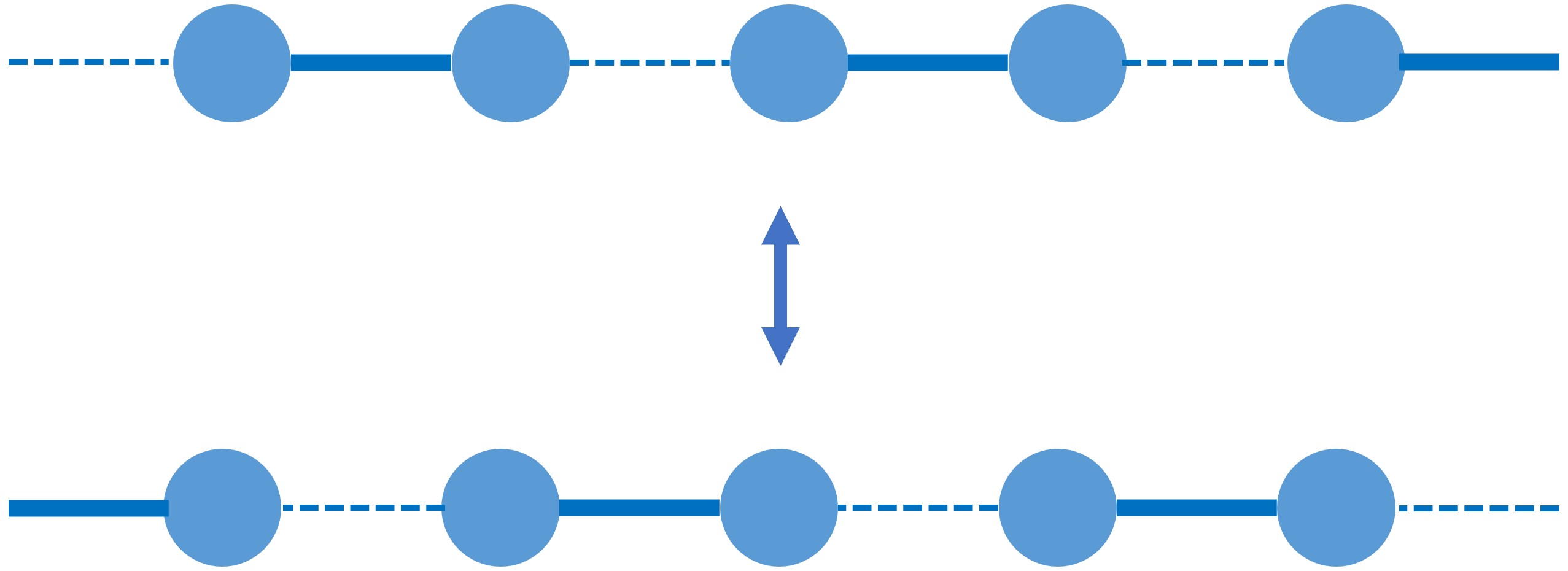}
\caption{\label{fig:RVB} Illustration of two dimerized states of a spin chain, where the solid and dashed lines represent strong and weak interactions between spins, respectively.}
\end{figure}

The resonating valence bond (RVB) state of quantum spin chains has received constant attention since Anderson described a copper oxide superconductor as such a state in 1987\cite{RVB}. The idea is based on the fact that adjacent spins in the lattice can form dimers in multiple ways. The term ``dimer" here refers to strong spin-spin correlation between adjacent spins. A system in an RVB state is considered to be resonating among all possible dimerized states. Fig. \ref{fig:RVB} is an illustration of a linear spin chain resonating between two dimerized states, in which strong spin-spin correlation occurs between different pairs of adjacent spins. The ground state of this chain is described by the average of these two states, so the interactions between each two adjacent spins are equal throughout the entire system. 

For a chain with a finite length, this translational symmetry is broken and spin dimerization occurs in the ground state. A measure of spin dimerization can be defined as the difference between two spin-spin correlation functions\cite{CFT}: 
\begin{equation}
    \Delta_N (i) = |\langle \textbf{S}_i \cdot \textbf{S}_{i+1} \rangle - \langle \textbf{S}_{i-1} \cdot \textbf{S}_{i} \rangle|,
\label{eq:DeltaNspinchain}
\end{equation}
in which $N$ is the number of sites/spins in the chain, and index $i$ denotes the $i$-th site in the lattice. This dimerization measure compares the strengths of interactions between two pairs of adjacent spins on the two sides of site $i$. The greater $\Delta_N$ is, the more dimerized the chain becomes. In conformal field theory (CFT), it has been shown that this measure decays as the number of sites grows in the chain following a power law\cite{CFT}, that is,
\begin{equation}
    \Delta_N (i) \propto {\left[ N \sin\left( \frac{\pi i}{N}  \right) \right]}^{-d},
\label{eq:dimerorderspinchain}
\end{equation}
where the exponent $d$ is a system-dependent parameter. For a spin-1/2 Heisenberg model, CFT predicts $d=1/2$. 

Motta et al. discovered a similar dimerized state of three-dimensional hydrogen chains, and concluded that the decay of the dimerization follows the same power-law order as spin chains do while the number of atoms in the chain increases\cite{Motta2020}. For a finite-size hydrogen chain with open boundary conditions, the dimerization results from the open ends of chains, which can be viewed as a local perturbation to the density. As the size of a chain grows, that perturbation has less and less influence on the atoms far away from the ends. The definition of a dimerization measure is borrowed from Eq.~(\ref{eq:DeltaNspinchain}): the difference between the strengths of two adjacent H-H bonds on both sides of the $i$-th atom. The strength of a bond is usually represented by the electron density at the bond critical points (BCP)\cite{BCP1,BCP2}, i.e., the local minima in electron density, and thus, the dimerization measure can be simply represented by the difference between two adjacent minima of the density on the two sides of the $i$-th atom:

\begin{equation}
\Delta_N (i) = \left|n^{i+1,\  i}_{\rm BCP} - n^{i,\  i-1}_{\rm BCP}\right|.
\label{eq:DeltaNBCP}
\end{equation}

\begin{figure}[t]
\centering
\includegraphics[scale=0.72]{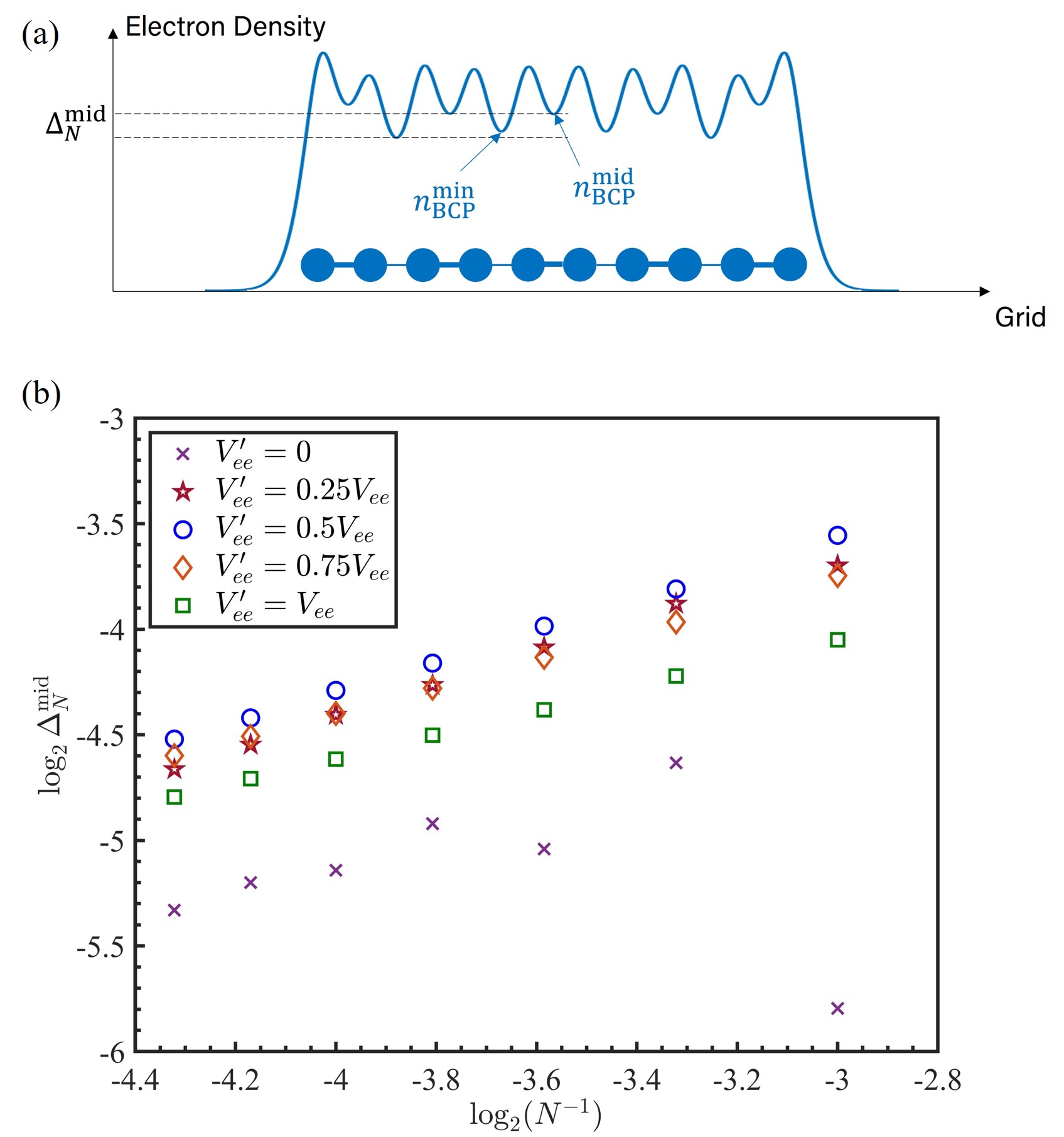}
\caption{(a) Dimerized electron density of ${\rm H_{10}}$, associated with the alternative presence of strong and weak bonds along the chain. The difference between the two dashed lines defines the dimerization measure $\Delta_N$. (b) Dependence of $\Delta^{\rm mid}_N$ on the size of hydrogen chains ($R = 3$a.u.) computed with DMRG. The electron-electron interaction $V_{ee}$ in DMRG is adjusted to study the effect of interaction strength on dimerization order.}
\label{fig:DeltaNdef}
\end{figure}

We now explore the existence of such a power-law order in 1D hydrogen chains with PDFT. For simplicity, we studied the density dimerization in the center of the chains. Choosing $i=N/2$ in Eq. (\ref{eq:DeltaNBCP}) gives

\begin{equation}
    \Delta^{\rm mid}_N = \Delta_N (N/2) = \left|n^{\rm mid}_{\rm BCP} - n^{\rm min}_{\rm BCP}\right|.
\end{equation}
For instance, as demonstrated in Fig.~\ref{fig:DeltaNdef}(a), $n^{\rm mid}_{\rm BCP}$ and $n^{\rm min}_{\rm BCP}$ are the two density minima on two different sides of the fifth atom in $\rm H_{10}$. One may conclude from Eq. (\ref{eq:dimerorderspinchain}) that 
\begin{equation}
    \Delta^{\rm mid}_N \propto N^{-d}
\label{eq:dimerorderHchain}
\end{equation}
if the dimerization in hydrogen chains also decays as a power-law order like spin chains.

Fig. \ref{fig:DeltaNdef}(b) plots the dimerization measure for a series of hydrogen chains with different sizes, calculated with DMRG at a fixed interatomic distance $R = 3$a.u. To elucidate how the strength of the electron-electron interaction ($V_{ee}$) influences the presence of dimerization, densities are calculated by tuning the strength of the electron-electron interaction. The effective interaction, $V_{ee}'$, is set to be in 0, 25\%, 50\%, 75\%, and 100\% of the strength of the original $V_{ee}$. The linearity of those logarithm plots indicates the validity of Eq. (\ref{eq:dimerorderHchain}). The magnitude of $\Delta_N$ represents the extent of the density dimerization in hydrogen chains, and the slope corresponds to the exponent $d$ in Eq. (\ref{eq:dimerorderHchain}), which measures the decay rate of the dimerization as the size of the chain increases. Table \ref{table:DeltaNSlope} lists the slopes and coefficients of determination for all data sets in Fig. \ref{fig:DeltaNdef}(b). When the electrons are non-interacting ($V_{ee}'=0$), the power law clearly breaks down. As $V_{ee}'$ increases from 0 to 100\% of $V_{ee}$, we observe an increase in the linearity, verifying that the presence of the power-law order in the decay of density dimerization is indeed a result of the electron-electron interaction. The linearity emerges even when the interaction is weak ($V_{ee}' = 0.25V_{ee}$). We also notice that different interaction strengths lead to different values for $d$. According to Table \ref{table:DeltaNSlope}, $d$ increases as $V_{ee}'$ increases from 0 to 0.5$V_{ee}$, where it reaches a maximum of $d\approx 0.724$, but further increasing the interaction strength above 50\% leads to a decrease in the value of $d$. For the fully interacting system, $d\approx 0.566$. 

\begin{table}[h]
	\caption{Slopes $d$ and coefficients of determination $\langle {\rm R}^2 \rangle$ of the plots in Fig. \ref{fig:DeltaNdef}(b).}
	\centering
    \setlength{\tabcolsep}{5mm}{
	\begin{tabular}{c c c}
        & & \\
		\hline\hline
		$V_{ee}'/V_{ee}$ & Slope ($d$) & $\langle {\rm R}^2 \rangle$  \\
   \hline
        0.00 & -0.101  & 0.0174\\
        0.25 & 0.686  & 0.9784 \\
        0.50 & 0.724 & 0.9997\\
        0.75 & 0.644 & 0.9997\\
        1.00 & 0.566 & 0.9998\\
		\hline
	\end{tabular}}
	\label{table:DeltaNSlope}
\end{table}

\begin{figure}[t]
\centering
\includegraphics[scale=0.72]{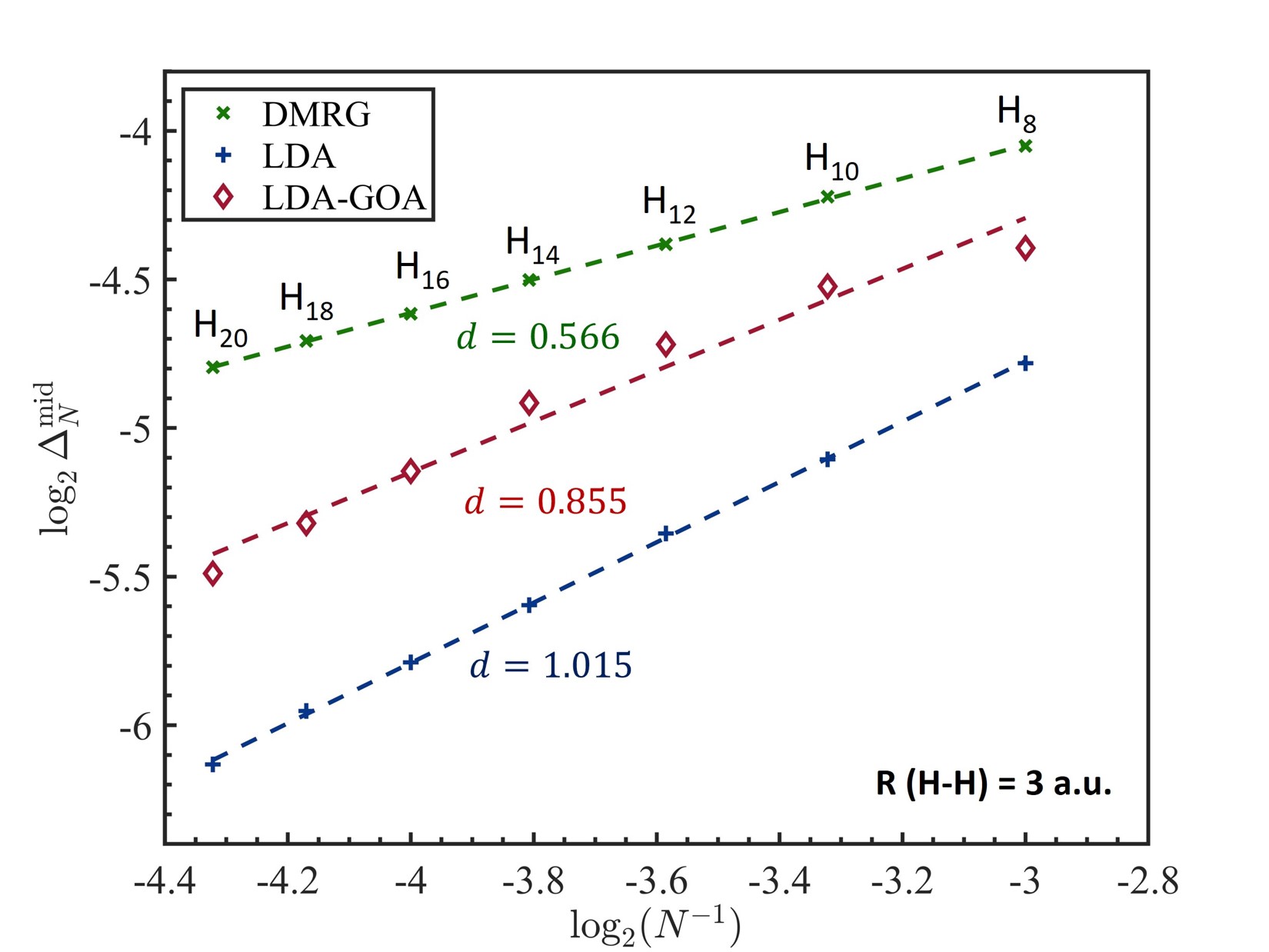}
\caption{\label{fig:Dimer_R3} Dependence of $\Delta^{\rm mid}_N$ on the size of hydrogen chains ($R = 3$a.u.). Results are calculated with DMRG (in green), LDA (in blue), and LDA-GOA (in red).}
\end{figure}

We now compare the density dimerization measure in hydrogen chains obtained {\em via} PDFT with and without the GOA. As is demonstrated in Fig. \ref{fig:Dimer_R3}, all of the three methods (DMRG, LDA, and LDA-GOA) produce linear $\log \Delta_N - \log (1/N)$ plots. According to Fig. \ref{fig:Dimer_R3}, not only does the LDA underestimate the magnitude of $\Delta_N$ for all the hydrogen chains listed above, but it also overestimates $d$ by $\sim$ 80\%. The underestimation of the magnitude of $\Delta_N$ by LDA implies that it generally underestimates how dimerized the electron density can be in hydrogen chains. 
Furthermore, the overestimation of the slope indicates that the LDA causes the dimerization to decay too rapidly as the hydrogen chain grows. By using the GOA in PDFT, {\em both} $\Delta_N$ and $d$ improve significantly over the LDA. The magnitudes of $\Delta_N$ are closer to DMRG results for all hydrogen chains, and the slope $d$ is predicted to be 0.855, a 30\% improvement. 

\begin{figure}[t]
\centering
\includegraphics[scale=0.72]{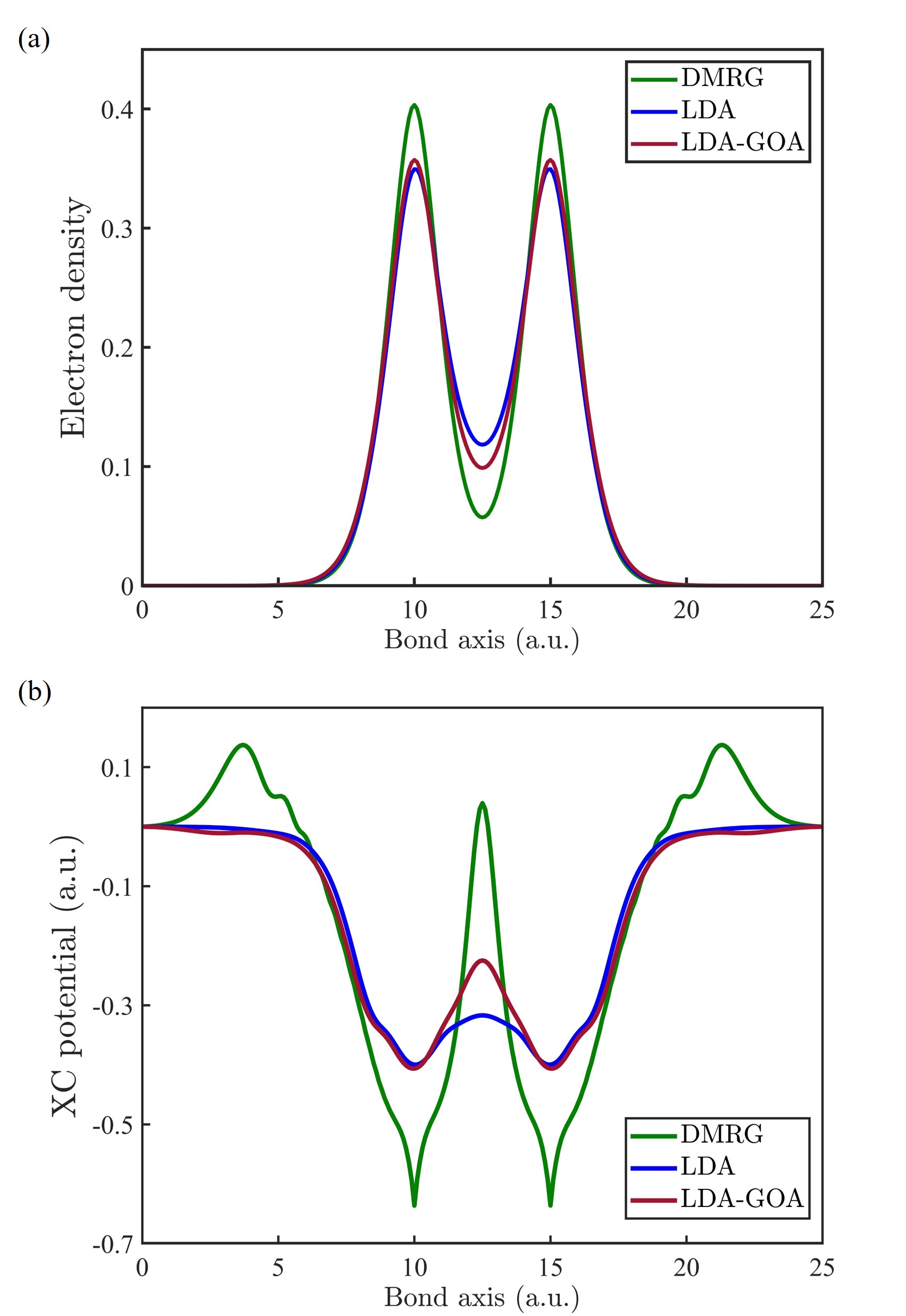}
\caption{\label{fig:H2n_v} (a) Ground-state density of $\rm H_{2}$ ($R=5$a.u.) computed from DMRG (in green), LDA (in blue), and LDA-GOA (in red). (b) The XC potential obtained from inverting the three densities in panel (a).} 
\end{figure}

This improvement can be mainly attributed to the density correction brought by the GOA, which can be 
seen also from the corresponding exchange-correlation potentials.
Here, we first use the simple example of $\rm H_2$ to demonstrate this correction. Fig. \ref{fig:H2n_v}(a) shows the ground-state density of a 1D hydrogen molecule ($\rm H_{2}$, $R=5$a.u.) computed with the three methods: DMRG, LDA, and LDA-GOA. Panel (b) of Fig. \ref{fig:H2n_v} compares the exchange-correlation potential $v_{\rm XC}(x)$ obtained from inverting the three densities in panel (a). Out of the three densities,  the (exact) DMRG density is the most localized around the nuclei, and the LDA density is the most delocalized; the GOA-LDA density falls in between. 
The XC potential inverted from the exact density (in green in panel (b)) exhibits sharp valleys around the nuclei and a high barrier at the BCP of the H-H bond, while the LDA potential only has two shallow valleys along with a much lower barrier in between the nuclei. This contrast in the potentials accounts for the differences in density localization seen in panel (a).  The 
GOA-LDA potential shows a higher barrier than the LDA at the BCP, whereas the valleys near the nuclei are almost the same as those in the LDA potential. 

\begin{figure}[t]
\centering
\includegraphics[scale=0.72]{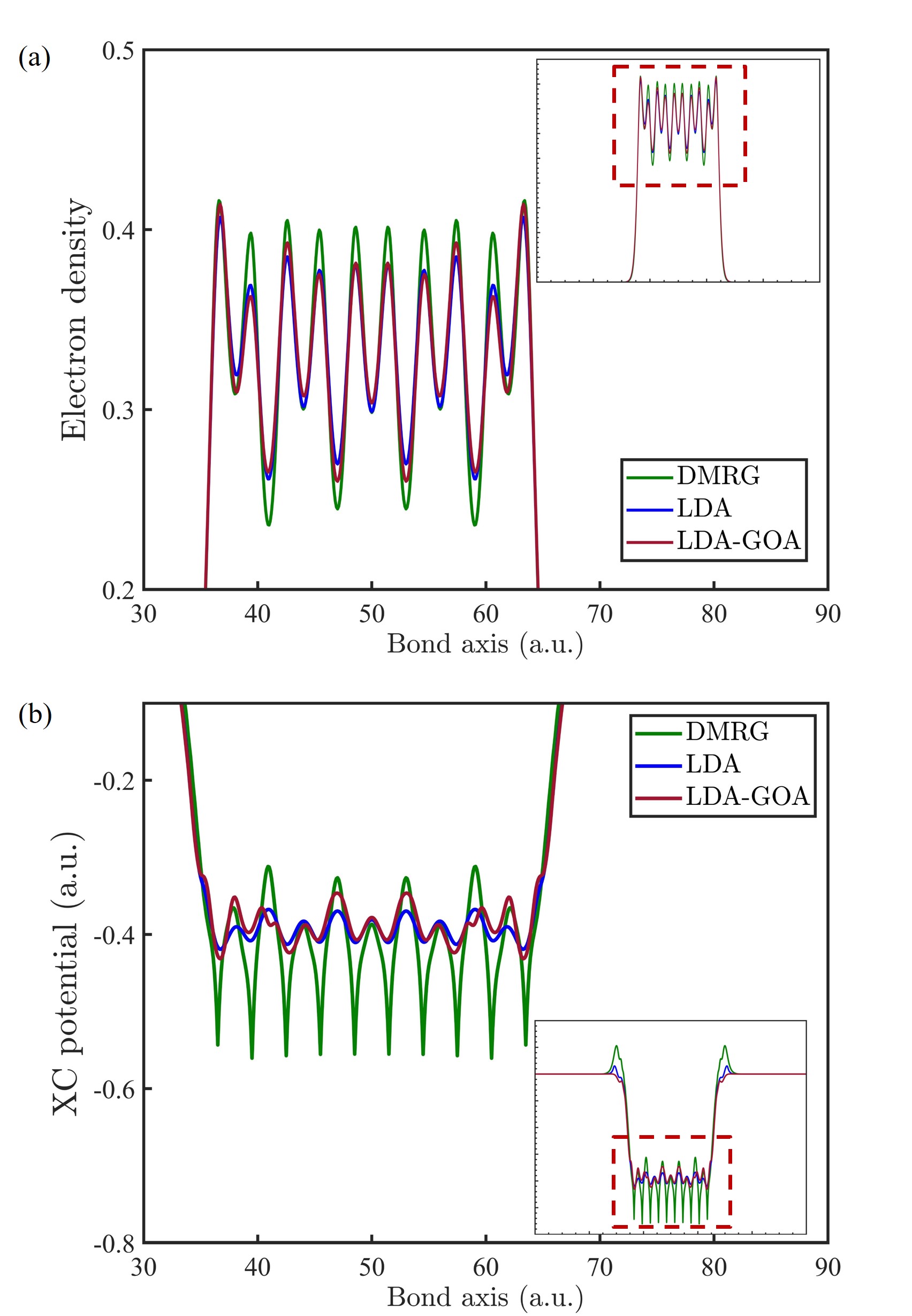}
\caption{\label{fig:H10n_v} (a) Ground-state density of $\rm H_{10}$ ($R=3$a.u.) computed from DMRG (in green), LDA (in blue), and LDA-GOA (in red). (b) The XC potential obtained from inverting the three densities in panel (a).}
\end{figure}

Similar corrections to the electron densities and the corresponding XC potentials occur in the case of the hydrogen chains. Unlike the case of $\rm H_2$, we are now more interested in the extent of density dimerization present in the chain. Fig. \ref{fig:H10n_v}(a) shows the ground-state density of $\rm H_{10}$ ($R=3$). DMRG produces a density in which electrons form stronger dimers than LDA. In other words, LDA underestimates the difference between strong and weak H-H bonds in the chain. Correspondingly, we see that the XC potential corresponding to the DMRG density displays four high barriers and five low barriers at the BCPs in panel (b) of Fig. \ref{fig:H10n_v}, while the heights of barriers in the LDA XC potential are considerably similar to each other. The GOA leads to a correction to the LDA density and produces results closer to those of DMRG. For most of the dimers (except for the two at the ends of the chain), the GOA drives the density to be more dimerized by increasing the heights of certain barriers in $v_{\rm XC}(x)$. Particularly, the two barriers near the center of the chain are close to those of the DMRG potential. However, the GOA does not improve the other two barriers of the LDA XC potential, leaving them almost unchanged. Clearly, a more elaborate approximation for the partition energy is needed to achieve this.

\clearpage
\section{\label{sec:exactdisso}Towards the Exact Dissociation Limit}
\begin{figure}[t]
\centering
\includegraphics[scale=0.7]{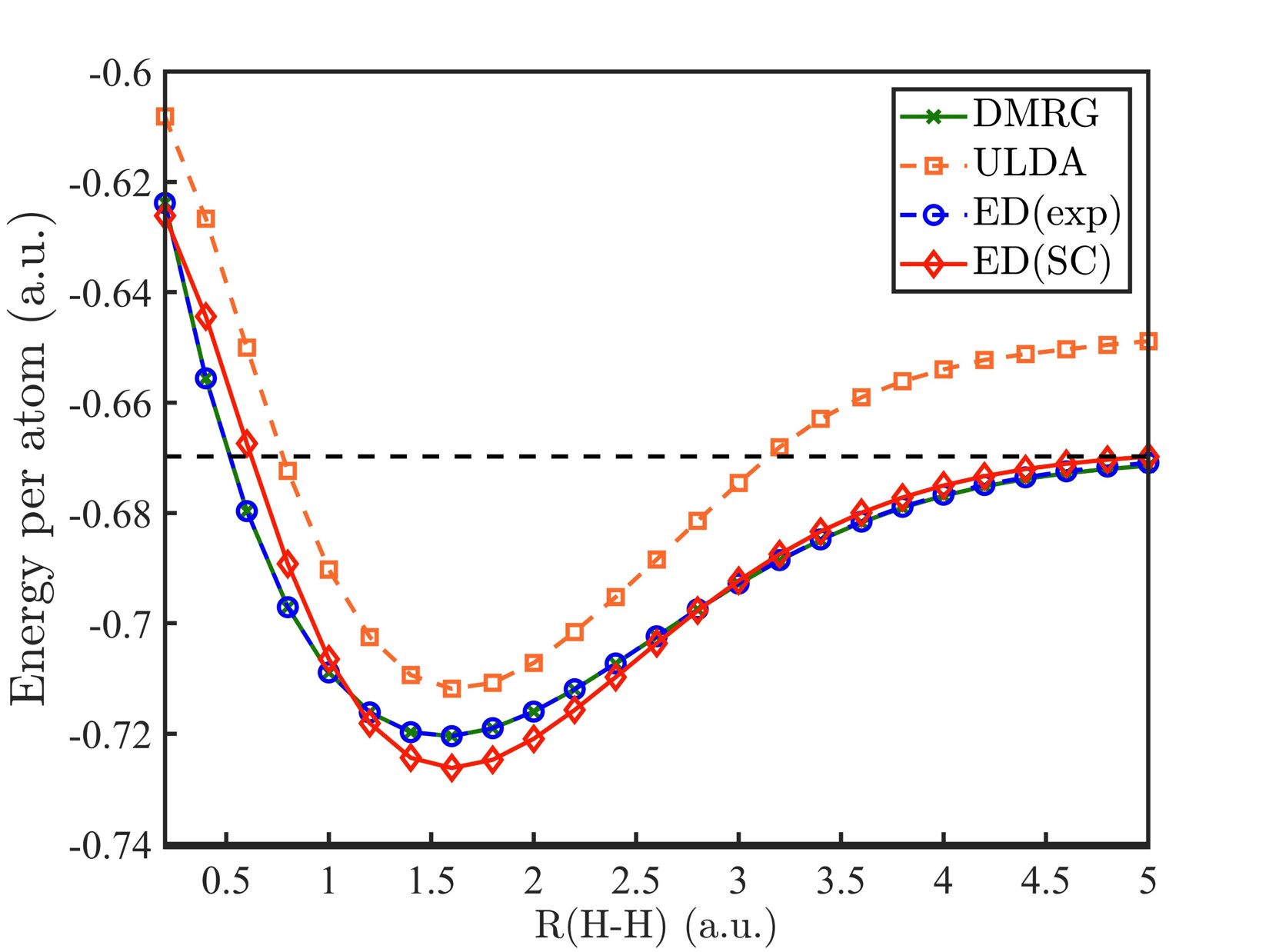}
\caption{\label{fig:H2ED} Dissociation curve of $\rm H_{2}$ computed with DMRG (green), ULDA (orange), and ED with SC (red) and exponential (blue) interactions. The dissociation limit ($E({\rm H}) = -0.670$a.u.) is marked by the dashed black line.}
\end{figure}

We showed in Sec. \ref{sec:dissociation} that the GOA yields energies similar to those of ULDA calculations (but without breaking the spin symmetry). We now compare with numerically exact DMRG calculations. Unlike 1D-DFT, DMRG adopts an exponential interaction in place of the soft Coulomb potential for $V_{NN}$ \cite{expDMRG}. To evaluate the effects of such a difference on our calculations, we perform two exact diagonalization (ED) calculations on 1D $\rm H_{2}$, using exponential (exp) and soft Coulomb (SC) interaction potentials. Fig. \ref{fig:H2ED} shows the dissociation curve of 1D $\rm H_{2}$ computed with DMRG (green), ULDA (orange), and ED with the SC (red) and exponential potential (blue). We see that the ED energies with the exponential potential match exactly those of DMRG. Replacing the exponential potential with the SC interaction in ED leads to a slightly different dissociation curve, but the differences between the two are generally distributed around the small-$R$ region around the equilibrium separation. For the large-$R$ region (when $R>3$), the SC potential does not have a significant impact on the energies. In contrast, the ULDA results (which are computed with an SC interaction), deviate significantly from the exact values for the entire range of separations.

\begin{figure}[t]
\centering
\includegraphics[scale=0.7]{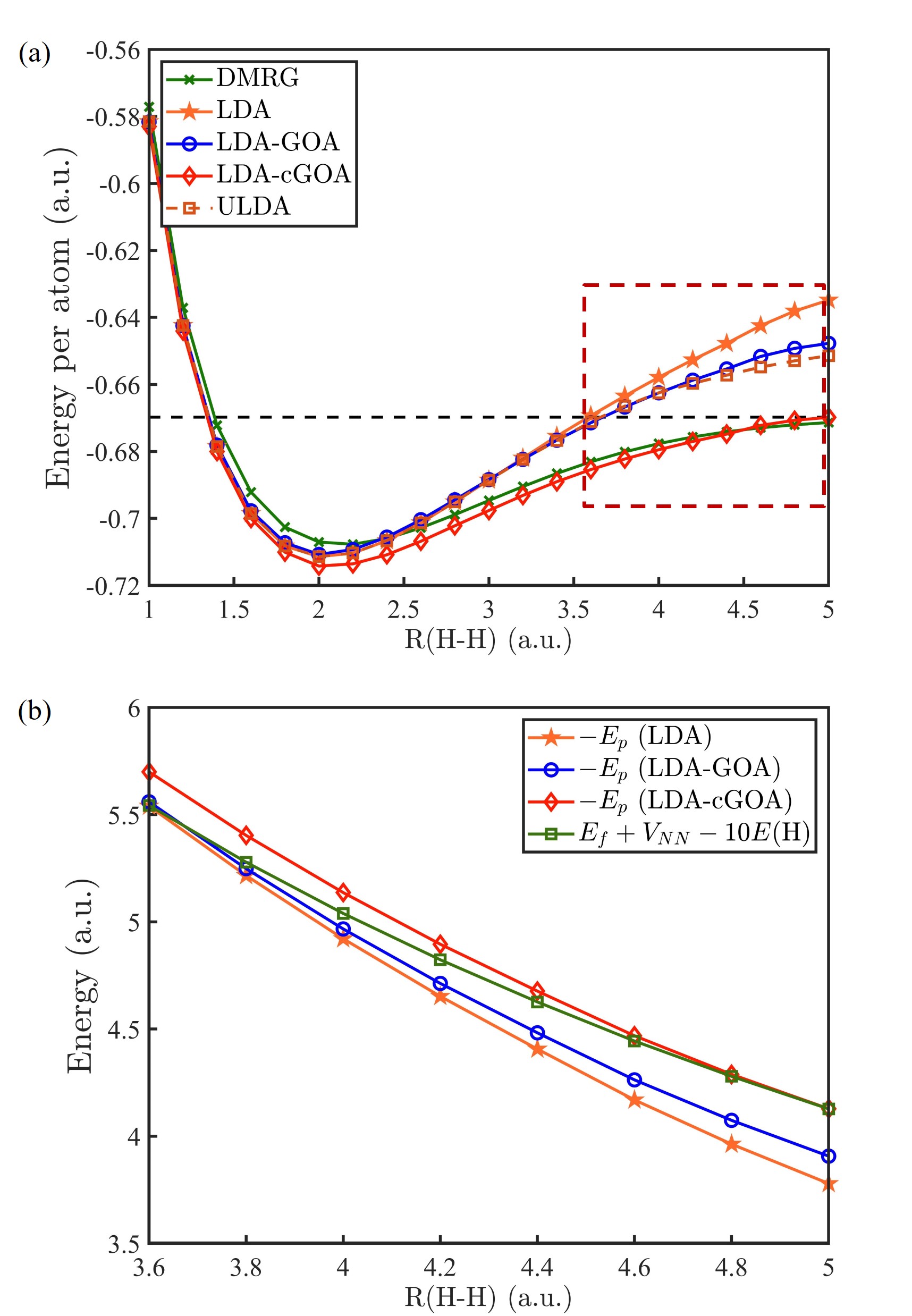}
\caption{\label{fig:H10_cGOA} (a) Dissociation curve of $\rm H_{10}$ obtained with DMRG (green), LDA (solid orange), ULDA (dashed orange), LDA-GOA (blue), and LDA-cGOA (red) calculations. The dissociation limit ($E({\rm H}) = -0.670$a.u.) is marked by the dashed black line. (b) $E_f+V_{NN}-10E$(H) compared with $-E_p$ calculated with LDA (orange), LDA-GOA (blue), and LDA-cGOA (red) in the large $R$ region. $E({\rm H}) = -0.670$a.u. is the exact ground-state energy of a hydrogen atom obtained with DMRG.}
\end{figure}

We now explore if PDFT can produce the correct dissociation limit $E(R \rightarrow \infty)$ of a hydrogen chain in 1D, improving over the LDA-GOA (which approaches the incorrect ULDA limit). The exact ground-state energy of a 1D hydrogen atom is predicted by DMRG to be -0.670a.u., while the 1D-ULDA yields -0.647a.u.\cite{LDAEX} As a result, there is a difference between ULDA and DMRG in $E(R \rightarrow \infty)$. 
To fix the error caused by 1D-ULDA in the dissociation limit, an additional correction to $E_p$ is made by modifying the non-additive Hartree term:
\begin{align}
    E_p^{\rm cGOA} = & \,T_s^{\rm nad} + E_{\rm ext}^{\rm nad}+S^{\rm GOA}E_{\rm XC}^{\rm nad}\notag\\
   & + \left( 1- \frac{1-S^{\rm GOA}}{N_f}\right) E_{\rm H}^{\rm nad},
    \label{eq:EpcGOA}
\end{align}
where the superscript $\rm cGOA$ stands for \textit{corrected generalized overlap approximation}. Fig. \ref{fig:H10_cGOA}(a) shows the dissociation curve of $\rm H_{10}$ calculated with PDFT using the cGOA for $E_p$, compared against the exact results obtained from DMRG, along with LDA and LDA-GOA. 
Focusing on the large-$R$ region ($R \in [3.6,5.0]$), we see that the LDA-cGOA calculation 
drives the curve to the correct dissociation limit, whereas all the other calculations (LDA, ULDA, and LDA-GOA) do not converge to the correct energy as $R \rightarrow \infty$.

We further check the cGOA by plotting the components of the total energy in PDFT. Fig. \ref{fig:H10_cGOA}(b) depicts $E_p$ and $E_f$ in the large $R$ region for $\rm H_{10}$, in which $-E_p$ is plotted for convenience. The three different $E_p$'s are extracted from LDA, LDA-GOA, and LDA-cGOA calculations. $E_f$ is the sum of fragment energies calculated using LDA for the XC energy, $V_{NN}$ is the soft Coulomb inter-nculear repulsion, and $E$(H) is the DMRG energy of a hydrogen atom. The total energy in PDFT reads $E_p(R)+E_f(R)+V_{NN}(R)$, so we consider PDFT to produce the correct dissociation limit of the system if $E_p(\infty)+E_f(\infty)+V_{NN}(\infty)-10E({\rm H}) \rightarrow 0$, which means that the magnitude of $-E_p$ and $E_f(R)+V_{NN}(R)-10E({\rm H})$ should match for large $R$'s in Fig. \ref{fig:H10_cGOA}(b). Now we look at the energies at $R=5$a.u. Although $R=5$a.u. is not yet the dissociation limit, 
we know from Fig. \ref{fig:H10_cGOA}(a) that $E_{\rm el}(R=5)+V_{NN}(R=5) \approx E_{\rm el}(\infty)$. In Fig. \ref{fig:H10_cGOA}(b), the difference between the LDA $-E_p$ (orange stars) and $E_f(R)+V_{NN}(R)-10E({\rm H})$ (green squares) at $R=5$a.u. can be identified as the static correlation error. LDA-GOA (blue circles) partly rectifies this error since that energy difference is smaller but remains non-zero at $R=5$a.u.  In contrast, $E_p$ obtained with LDA-cGOA (red diamonds) exactly cancels off $E_f(R)+V_{NN}(R)-10E({\rm H})$ at $R=5$ and yields the correct dissociation limit of $\rm H_{10}$.

\section{\label{sec:conclusion}Conclusion}
The overlap approximation of PDFT \cite{OA} has been generalized to be applicable to multi-fragment systems (GOA). When applied 
 to strongly-correlated 1D hydrogen chains,
we observed that using the LDA for the fragments, 
PDFT with the GOA produces dissociation curves that are close to those of 1D unrestricted-LDA calculations while retaining the correct spin symmetries. Furthermore, the GOA improves upon the LDA electron densities of hydrogen chains leading to dimerization measures that approach those of DMRG, partially capturing an important signature of strongly-correlated physics.  Moreover, with an additional correction (cGOA) to the non-additive Hartree component to make the partition energy more negative as needed, LDA-cGOA calculations in PDFT produce the correct dissociation limit for $\rm H_{10}$. 

We note that the density correction induced by the GOA addresses only part of the error in the LDA density, so it is insufficient to describe the full density dimerization observed from DMRG calculations. Nevertheless, the GOA represents a step in the right direction.   Looking ahead, it will be important to: (1)  Find more robust and generally applicable approximations to the partition energy of PDFT, perhaps based on the foundation of the GOA; and (2) Extend these calculations to 3D systems so that PDFT can be applied to more realistic strongly-correlated materials.

The method described in this work does not abandon the use of standard functionals for the fragments (here, the LDA), suggesting that a promising density-functional route to strongly-correlated physics consists on supplementing Perdew's Jacob's ladder of approximations for the {\em fragments} with a smaller ladder of approximations for the {\em inter-fragment} interactions.  After all, as is well known, to get up to heaven one needs a big ladder: A big ladder and another smaller one \cite{LaBamba}.

{\bf {\em Acknowledgement:}} 
This work was supported by the U.S. National Science Foundation under Grant No. CHE-2306011. Acknowledgement is also made to the donors of the American Chemical Society Petroleum Research Fund for partial support of this research under grant No. 62544-ND6.

{\bf {\em Code Availability:}} 
The MATLAB code used for all the 1D-PDFT calculations reported here is available from https://github.com/yishi62751/PDFT\_1D.git.

\bibliographystyle{unsrt}  






\nocite{*}
\bibliography{references}

\end{document}